# A structural theory for a 3D isotropic linear-elastic finite body

E Hanukah*, Bella Goldshtein**

*Faculty of Mechanical Engineering
Technion - Israel Institute of Technology
Haifa, 32000, Israel
**Department of Fine Arts
University of Haifa
Mt. Carmel, Haifa 31905, Israel
Email: eliezerh@tx.technion.ac.il

## Abstract

The development of a nonlinear structural theory (model) for isotropic linear-elastic finite continua is the main objective of the study. To derive the theory, we used Taylor's multivariable expansion and Bubnov-Galerkin's weak formulation. The resulting formulation is consistent with elasticity theory. As a special case, we consider second order theory, resulting in 30 internal degrees of freedom (IDF). A set of 30 ordinary differential equations (ODE) is derived and must be solved to determine the deformation field. ODEs depend on the initial and actual geometry of the structure, loads and material properties. Our theory can be used to derive analytical or numerical solutions. Aspects of generalization of the method to solids with nonlinear constitutive relations are presented.

## 1. Introduction

An elastic body with a finite size has an infinite number of material points; their motion is governed by partial differential equations. For most engineering applications, an analytical solution is not available. Governed by Newton–Euler equations of motion, a rigid body has six degrees of freedom; however, for a deformable body this is not sufficient. Several numerical approaches like finite element or finite differences are widely used to achieve numerical solutions; however, analytical solutions cannot be produced using these methods.

To the best of our knowledge there are two main approaches used to formulate closed-form theories for elastic 3D bodies: pseudo-rigid bodies and the theory of Cosserat point or Cosserat-like theories (see for example [1-2]). It seems both methods have the same origins (see [3-9]). Analytical solutions can be produced (see for example [10-13]). Both methods can be converted to the finite element method (see [14-17]).

To the best of our knowledge, the pseudo-rigid body method is mostly used to investigate dynamics of bodies which, in addition to rigid body motion, can undergo homogeneous deformations for multi-body dynamic simulations which combine deformable solids with objects modeled as rigid bodies (see for example [18]). Papadopoulos [19] has developed a second order theory of a pseudo-rigid body which has 30 degrees of freedom. It seems that the establishment of a pseudo-rigid body model on the basis of continuum mechanics is a delicate and unresolved issue (see Steigmann [20], Casey [21], Cohen and Muncaster [1]).

Among other applications, the theory of Cosserat point is used to formulate finite elements (CPE) for nonlinear elasticity (see for example Rubin [2], Nadler and Rubin [22], Loehnert et al. [23], Boerner et al. [24], Jabareen and Rubin [25-26], Jabareen et al [27]). Constitutive equations are derived using closed-form energy function. The energy is separated into two parts: the first uses the average measures of deformation and the second is calibrated to satisfy analytical solutions from linearized linear elasticity. In some elements, special treatment of volume changes is necessary (see [26, 27]). Locking and hour glassing does not occur. For undistorted elements, superior behavior is presented; however, every new formulation needs to be "calibrated," a process requiring great expertise and creativity.

In the present study we wish to present a method to formulate closed-form structural models for isotropic linear-elastic finite bodies. We confine ourselves to second order approximation; however, the generalization is immediate. We put the emphasis on consistency with standard continuum mechanics, minimum ad-hoc assumptions, and a unified approach for all the structures. Basic kinematic approximation is polynomial multivariable expansion of the unknown deformation field. The expansion leads to ten unknown vector quantities and, consequently, to 30 scalar unknowns. We define 30 internal degrees of freedom (IDF) such that they have zero value for the initial configuration (initial configuration is prescribed, no freedom allowed). With the help of basic kinematic approximation and weak Bubnov-Galerkin formulation, 30 ordinary differential equations (ODE) are derived, and have to be solved to uniquely determine the deformation field. Geometrically nonlinear equations, fully coupled, depend on material constants, loads, initial and actual geometries. This approach can be applied to analytical investigation of structural behavior, can produce analytical or numerical solutions, or can be converted to finite elements using CPE's conversion technique (see for example [27]). Possible generalization to nonlinear materials is suggested.

An outline of the paper is as follows. Section 2 presents the derivation of the basic kinematic approximation and derivation of governing equations of the theory, including a very brief insight regarding linearization and some very basic guidelines for material generalization.

Section 3 presents the summary and conclusions. Finally, Appendices A record details of the developments.

## 2. Method.

The initial configuration of the structure (body) is characterized by the center of mass denoted by $\mathbf{X}_0$ , three covariant base vectors $\{\mathbf{G}_1, \mathbf{G}_2, \mathbf{G}_3\}$ (which characterize the initial geometry (see for example [27])), and curvilinear material coordinates $\{\theta^1, \theta^2, \theta^3\}$ which are zero at the $\mathbf{X}_0$ . Material point $\mathrm{X}$ of the structure can be exactly represented by

$$\mathbf{X} = \mathbf{X}_0 + \theta^1 \mathbf{G}_1 + \theta^2 \mathbf{G}_2 + \theta^3 \mathbf{G}_3 \tag{1}$$

The actual position $\mathbf{x}$ of $\mathrm{X}$ is given by $\mathbf{x} = \boldsymbol{\chi}(\mathbf{X}(\theta^1, \theta^2, \theta^3), t)$ , where $\boldsymbol{\chi}$ is motion to be determined. We use Taylor's multivariable expansion up to second order about the center of mass to derive basic kinematic approximation

$$\begin{aligned}
\mathbf{x}^{h} = {} & \boldsymbol{\chi}(\mathbf{x}_0, t) + \theta^1 \frac{\partial \boldsymbol{\chi}(\mathbf{x}_0, t)}{\partial \theta^1} + \theta^2 \frac{\partial \boldsymbol{\chi}(\mathbf{x}_0, t)}{\partial \theta^2} + \theta^3 \frac{\partial \boldsymbol{\chi}(\mathbf{x}_0, t)}{\partial \theta^3} + \\
& + \theta^1 \theta^2 \frac{\partial^2 \boldsymbol{\chi}(\mathbf{x}_0, t)}{\partial \theta^1 \partial \theta^2} + \theta^1 \theta^3 \frac{\partial^2 \boldsymbol{\chi}(\mathbf{x}_0, t)}{\partial \theta^1 \partial \theta^3} + \theta^2 \theta^3 \frac{\partial^2 \boldsymbol{\chi}(\mathbf{x}_0, t)}{\partial \theta^2 \partial \theta^3} + \\
& + \theta^1 \theta^1 \frac{\partial^2 \boldsymbol{\chi}(\mathbf{x}_0, t)}{2 \partial \theta^1 \partial \theta^1} + \theta^2 \theta^2 \frac{\partial^2 \boldsymbol{\chi}(\mathbf{x}_0, t)}{2 \partial \theta^2 \partial \theta^2} + \theta^3 \theta^3 \frac{\partial^2 \boldsymbol{\chi}(\mathbf{x}_0, t)}{2 \partial \theta^3 \partial \theta^3}
\end{aligned} \tag{2}$$

where $\mathbf{x}_0$ is the actual position of the center of mass. Next we define the shape functions $\mathrm{N}^{\mathrm{i}}$ by (3a) and vector kinematic variables $\mathbf{x}_{\mathrm{i}}$ by (3b)

$$\begin{gathered}
\mathrm{N}^0 = 1 \;,\; \mathrm{N}^1 = \theta^1 \;,\; \mathrm{N}^2 = \theta^2 \;,\; \mathrm{N}^3 = \theta^3 \;,\; \mathrm{N}^4 = \theta^1 \theta^2 \\
\mathrm{N}^5 = \theta^1 \theta^3 \;,\; \mathrm{N}^6 = \theta^2 \theta^3 \;,\; \mathrm{N}^7 = \theta^1 \theta^1 \;,\; \mathrm{N}^8 = \theta^2 \theta^2 \;,\; \mathrm{N}^9 = \theta^3 \theta^3
\end{gathered} \tag{3a}$$

$$
\begin{gathered}
\mathbf{x}_0 = \boldsymbol{\chi}(\mathbf{x}_0, t) \\
\mathbf{x}_1 = \frac{\partial \boldsymbol{\chi}(\mathbf{x}_0, t)}{\partial \theta^1} \;,\; \mathbf{x}_2 = \frac{\partial \boldsymbol{\chi}(\mathbf{x}_0, t)}{\partial \theta^2} \;,\; \mathbf{x}_3 = \frac{\partial \boldsymbol{\chi}(\mathbf{x}_0, t)}{\partial \theta^3} \\
\mathbf{x}_4 = \frac{\partial^2 \boldsymbol{\chi}(\mathbf{x}_0, t)}{\partial \theta^1 \partial \theta^2} \;,\; \mathbf{x}_5 = \frac{\partial^2 \boldsymbol{\chi}(\mathbf{x}_0, t)}{\partial \theta^1 \partial \theta^3} \;,\; \mathbf{x}_6 = \frac{\partial^2 \boldsymbol{\chi}(\mathbf{x}_0, t)}{\partial \theta^2 \partial \theta^3} \\
\mathbf{x}_7 = \frac{\partial^2 \boldsymbol{\chi}(\mathbf{x}_0, t)}{2\partial \theta^1 \partial \theta^1} \;,\; \mathbf{x}_8 = \frac{\partial^2 \boldsymbol{\chi}(\mathbf{x}_0, t)}{2\partial \theta^2 \partial \theta^2} \;,\; \mathbf{x}_9 = \frac{\partial^2 \boldsymbol{\chi}(\mathbf{x}_0, t)}{2\partial \theta^3 \partial \theta^3}
\end{gathered} \tag{3b}
$$

Using (3a) and (3b), the basic kinematic approximation (2) can be conveniently expressed by

$$
\mathbf{x}^h = \sum_{j=0}^{9} N^j(\theta^m)\mathbf{x}_j(t) \;,\; (m = 1,2,3) \tag{4}
$$

(4) has to admit initial representation (1) so that initial values $\mathbf{X}_i$ of kinematic variables $\mathbf{x}_i$ are determined by (5a) and representation (1) can be expressed in terms of vector kinematic variables (5b)

$$
\begin{gathered}
\mathbf{x}^h \to \mathbf{X} \;,\; \mathbf{x}_i \to \mathbf{X}_i \;,\; (i = 0,..,9) \\
\mathbf{X}_1 = \mathbf{G}_1 \;,\; \mathbf{X}_2 = \mathbf{G}_2 \;,\; \mathbf{X}_3 = \mathbf{G}_3 \\
\mathbf{X}_j = \mathbf{0} \;,\; (j = 4,..,9)
\end{gathered} \tag{5a}
$$

$$
\mathbf{X} = \sum_{i=0}^{9} N^i(\theta^m)\mathbf{X}_i \;,\; (m = 1,2,3) \tag{5b}
$$

Second order approximation leads to ten vector kinematic variables, which has 30 components. Ten vectorial equations and consequently thirty scalar equations have to be derived to determine the deformation field. Firstly, we conveniently define internal degrees of freedom such that all the IDFs vanish for initial configuration.

$$
\mathbf{x}_i = \sum_{k=1}^{3} \hat{x}_i^{\,k}(t)\mathbf{G}_k \;,\; \hat{x}_i^{\,k}(t) = b^{3i+k}(t) + \mathbf{X}_i \bullet \mathbf{G}^k \;,\; (i = 0,..,9, k = 1,2,3) \tag{6}
$$

where $b^i(t)$ stands for IDFs, and it is noticed that $\left(\mathbf{X}_i \bullet \mathbf{G}^k\right)_{i=4,..,9} = 0$. In the present study, we used the second order expansion on $\mathbf{x}$ leading to 10 shape functions, 10 kinematic variables and 30 degrees of freedom. For higher order theory, additional terms from polynomial expansion should be used leading to additional shape functions, vector kinematic variables, and IDFs, but the manner by which generalization to higher order is done remains unchanged (steps (2)-(6)).

Conservation of mass with balance of linear momentum in actual configuration is given by (7a) and constitutive equation of linear elasticity by (7b)

$$\begin{gathered} \rho \dot{\mathbf{v}} = \rho \mathbf{b} + \mathrm{div}(\boldsymbol{\sigma}) \\ \boldsymbol{\sigma} = \lambda (\mathbf{E} \bullet \mathbf{I}) \mathbf{I} + 2\mu \mathbf{E} \ , \ \boldsymbol{\sigma} = \boldsymbol{\sigma}^T \end{gathered} \tag{7a, b}$$

where $\rho$ denote mass density, $\mathbf{v}$ denotes the velocity of material particle, a superposed dot denotes time differentiation, $\mathbf{b}$ denotes body force per unit of mass, $\lambda$ and $\mu$ are Lame constants, $\boldsymbol{\sigma}$ is a symmetric Cauchy stress tensor, $\mathrm{div}(\bullet)$ denotes the divergence operator $\mathrm{div}(\bullet) = (\bullet)_{,j} \mathbf{g}^j$, $\mathbf{I}$ denotes second order identity tensor, and $\mathbf{E}$ denotes Cauchy's strain tensor.

Basic kinematic approximation (4) is used to approximate $\dot{\mathbf{v}}$ , $\mathbf{E}$ and $\boldsymbol{\sigma}$ (see appendix A). Insertion of these approximations to the balance of linear momentum leads to

$$\mathbf{R} = \mathrm{div}(\boldsymbol{\sigma}^h) + \rho \mathbf{b} - \rho \dot{\mathbf{v}}^h \tag{8a}$$

The residual $\mathbf{R}$ is usually not zero since an approximation is made. The residual will be reduced to zero in a weak sense by multiplying the residual by shape functions (3a) and by integrating over the whole body (see for example [30]). Partial integration, divergence theorem and introduction of the traction boundary condition $\mathbf{t}$ lead to

$$\int_{\chi(\Omega_0)} N^i \rho \dot{\mathbf{v}}^h dv = \int_{\chi(\Omega_0)} N^i \rho \mathbf{b} dv + \int_{\chi(\partial\Omega_0)} N^i \mathbf{t} da - \int_{\chi(\Omega_0)} \boldsymbol{\sigma}^h (N^i{}_{,m} \mathbf{g}^m) dv \quad (i=0,..,9) \tag{8b}$$

Note that (8b) is a set of ten vector equations. We define convenient kinematic quantities $\mathbf{B}^i$ by

$$\mathbf{B}^i = \sum_{m=1}^{3} N^i{}_{,m} (g^{1/2} \mathbf{g}^m) \ , \ (i=0,..,9) \tag{9}$$

with the help of the next definitions

$$\begin{aligned} M^{ij} &= \int_{\chi(\Omega_0)} \rho N^i N^j dv \ , \ (i,j=0,..,9) \\ \mathbf{f}^i_{body} &= \int_{\chi(\Omega_0)} N^i \rho \mathbf{b} dv \ , \ (i=0,..,9) \\ \mathbf{f}^i_{ext} &= \int_{\chi(\partial\Omega_0)} N^i \mathbf{t} da \ , \ (i=0,..,9) \\ \mathbf{f}^i_{int} &= \int_{\chi(\Omega_0)} \mathbf{B}^i \boldsymbol{\sigma}^h \frac{1}{g^{1/2}} dv \ , \ (i=0,..,9) \end{aligned} \tag{10a,b,c,d}$$

definitions (9),(10a,b,c,d) and approximation (4) is used to rewrite (8b) as

$$\sum_{j=0}^{9} M^{ij} \dot{\mathbf{v}}_j = \mathbf{f}^i_{body} + \mathbf{f}^i_{ext} - \mathbf{f}^i_{int} \ , \ (i=0,..,9) \ , \tag{11}$$

where $\mathbf{v}_j = \dot{\mathbf{x}}_j$. (11) provides 10 vectorial equations for 10 vector kinematic variables $\mathbf{x}_j$. For $i=0$ equations (11) produce global balance of linear momentum.

The important question is: Can internal forces $\mathbf{f}^i_{int}$ be computed using symbolical software? In the present study, we used Maple to derive the detailed equations. We recall that $dv / g^{1/2} = d\theta^1 d\theta^2 d\theta^3$; the basic guideline is that monomials of coordinates $(\theta^1)^m (\theta^2)^n (\theta^3)^k$ where $m,n,k = 1,2,3$ can be integrated. Kinematic quantities $\mathbf{B}^i$ depend on monomials of the coordinates. Consequently, if $\boldsymbol{\sigma}^{h}$ includes only monomial terms with respect to coordinates, then symbolic integration can be performed. For linear elasticity $\boldsymbol{\sigma}^{h}$ is given by (A.5); approximated Caushy stress consists of geometry dependent terms, time dependent IDFs and coordinate monomials. For isotropic linear-elastic body symbolic integration $\mathbf{f}_{int}^{i} = \int_{\chi(\Omega_0)} \mathbf{B}^{i} \boldsymbol{\sigma}^{h} \frac{1}{g^{1/2}} dv$ can be performed

How can integrals for $\mathbf{f}_{int}^{i}$ be evaluated for nonlinear constitutive law, which will lead to rational or logarithmic terms, etc. with respect to coordinates? One can suggest numerical integration (the standard finite element method uses numerical integration to evaluate integrals), but then the resulting theory will not be able to produce analytical solutions. The other suggestion might be to use energy formulation to formulate constitutive laws for $\mathbf{f}_{int}^{i}$ . Cosserat point element theory uses structural energy formulation to derive $\mathbf{f}_{int}^{i}$; however, it is mostly used for nonlinear elasticity and to the best of our knowledge, there is no one unified procedure of energy formulation that is good for all structures and all the approximation orders. We suggest polynomial multivariable expansion of $\boldsymbol{\sigma}^{h}$ about the center of mass. What order of expansion should be used? Authors emphasize that in some sense this is an ad-hoc discussion, and quantitative investigation is needed. However, first order is inadequate for expressing nonlinearity; therefore, second order might be used, if found to provide accurate predictions. It is noted that for linear elasticity, $\boldsymbol{\sigma}^{h}$ is up to third order with respect to coordinates. If $\boldsymbol{\sigma}^{h}$ of nonlinear material will be expanded to third order, then the resulting $\mathbf{f}_{int}^{i}$ expressions will be no longer physically (meaning number of terms in the expressions) than expressions in linear elasticity. Authors emphasize that in a sense, discussion regarding the order of approximation of $\boldsymbol{\sigma}^h$ is ad-hoc and hand-waving; however, it does present some guidelines for further material generalization.

Using (11) we define

$$\mathbf{R}^i = \sum_{j=0}^{9} M^{ij}\dot{\mathbf{v}}_j - \mathbf{f}^i_{body} - \mathbf{f}^i_{ext} + \mathbf{f}^i_{int} \ , \ (i = 0,..,9) \qquad (12a)$$

using (12a) and (6) structural scalar equation for internal degrees of freedom defined

$$Eq_{3i+k}(t) = \mathbf{R}^i \bullet \mathbf{G}^k \ , \ (i = 0,..,9),(k = 1,2,3) \qquad (13)$$

Equations (13) are non-linear coupled expressions that depend on geometric terms, internal degrees of freedom, material properties and tractions applied on the body. Equations (13) are a set of 30 ODEs with respect to 30 IDFs, which must be solved to uniquely determine approximated deformation field (4).

Authors emphasize that for some specific geometry and loads these equations may be solved analytically. For other cases, numerical schemes can be applied (see for example [30]). Numerical schemes require use of stiffness matrix. Some advantages of closed-form stiffness matrixes are presented in [28-29]. The closed-form stiffness matrix of the system (13) is

$$\left\{ \begin{matrix} Eq_1(b^1(t),..,b^{30}(t)) = 0 \\ ... \\ ... \\ Eq_{30}(b^1(t),..,b^{30}(t)) = 0 \end{matrix} \right\} \ , \ K_{ij} = \frac{\partial Eq_i}{\partial b^j} \ , \ (i,j = 0,..,9) \qquad (14a,b)$$

## 3. Conclusions

In this study we developed a framework for structural modeling of isotropic linear-elastic finite bodies. We presented a systematic derivation of basic kinematic approximation for motion, resulting in a definition of shape functions and the notion of number of degrees of freedom. A convenient definition of internal degrees of freedom was introduced. Basic kinematic approximation and Bubnov-Galerkin weak form was used to systematically formulate an ordinary differential equation for every IDF. ODEs are nonlinear and coupled, and they depend on the initial and actual configurations of finite body, material properties and loads. The tangent stiffness matrix, if needed, is derived in a straightforward manner by differentiating 30 ODEs with respect to 30 IDFs. Generalization of the methodology to nonlinear materials is suggested.

**Acknowledgment**

The authors greatly acknowledge Ms. Tamara Sharbatova and Mr. Oren Hanuka for financial support of the research. The authors acknowledge Dr. Alexander Goldshtein, whose healthy criticism led us to deeper insights. The authors wish to acknowledge Mr. Levy Hanuka and Mrs. Ora Goldshtein for their interest and support.

**Appendix A:**

Covariant base vectors and their reciprocal are defined by

$$\begin{gathered}\mathbf{g}_i = \mathbf{x}^h{}_{,i} \\ \mathbf{g}^1 = \frac{\mathbf{g}_2 \times \mathbf{g}_3}{g^{1/2}} \ , \ \mathbf{g}^2 = \frac{\mathbf{g}_3 \times \mathbf{g}_1}{g^{1/2}} \ , \ \mathbf{g}^3 = \frac{\mathbf{g}_1 \times \mathbf{g}_2}{g^{1/2}} \ , \ \mathbf{g}_i \bullet \mathbf{g}^j = \delta_i^{\ j} \\ \mathbf{G}^1 = \frac{\mathbf{G}_2 \times \mathbf{G}_3}{G^{1/2}} \ , \ \mathbf{G}^2 = \frac{\mathbf{G}_3 \times \mathbf{G}_1}{G^{1/2}} \ , \ \mathbf{G}^3 = \frac{\mathbf{G}_1 \times \mathbf{G}_2}{G^{1/2}} \\ G_{ij} = \mathbf{G}_i \bullet \mathbf{G}_j \ , \ G^{ij} = \mathbf{G}^j \bullet \mathbf{G}^j \end{gathered} \tag{A.1}$$

where the commas denote partial differentiation with respect to $\theta^i$. Using (4) and definition (5) components of base vectors become

$$\begin{gathered}\mathbf{g}_i = \sum_{k=1}^{3} \hat{g}_i^{\ k} \mathbf{G}_k \ , \ \hat{g}_i^{\ k} = \sum_{j=1}^{10} N^j{}_{,i} \, \hat{x}_j^{\ k} \ , \ (i=1,2,3) \\ \mathbf{g}^i = \frac{1}{g^{1/2}} \sum_{k=1}^{3} \bar{g}^i{}_k \mathbf{G}^k \ , \ (i=1,2,3) \end{gathered} \tag{A.2a}$$

where components $\bar{g}^i{}_k$ and $g^{1/2}$ are detailed in terms of $\hat{g}_i^{\ k}$ components in (A.2b)

$$\begin{gathered}\bar{g}^1{}_1 = G^{1/2}\left(\hat{g}_2^{\ 2}\hat{g}_3^{\ 3} - \hat{g}_2^{\ 3}\hat{g}_3^{\ 2}\right) \ , \ \bar{g}^1{}_2 = G^{1/2}\left(\hat{g}_2^{\ 3}\hat{g}_3^{\ 1} - \hat{g}_2^{\ 1}\hat{g}_3^{\ 3}\right) \\ \bar{g}^1{}_3 = G^{1/2}\left(\hat{g}_2^{\ 1}\hat{g}_3^{\ 2} - \hat{g}_2^{\ 2}\hat{g}_3^{\ 1}\right) \ , \ \bar{g}^2{}_1 = G^{1/2}\left(\hat{g}_3^{\ 2}\hat{g}_1^{\ 3} - \hat{g}_3^{\ 3}\hat{g}_1^{\ 2}\right) \\ \bar{g}^2{}_2 = G^{1/2}\left(\hat{g}_3^{\ 3}\hat{g}_1^{\ 1} - \hat{g}_3^{\ 1}\hat{g}_1^{\ 3}\right) \ , \ \bar{g}^2{}_3 = G^{1/2}\left(\hat{g}_3^{\ 1}\hat{g}_1^{\ 2} - \hat{g}_3^{\ 2}\hat{g}_1^{\ 1}\right) \\ \bar{g}^3{}_1 = G^{1/2}\left(\hat{g}_1^{\ 2}\hat{g}_2^{\ 3} - \hat{g}_1^{\ 3}\hat{g}_2^{\ 2}\right) \ , \ \bar{g}^3{}_2 = G^{1/2}\left(\hat{g}_1^{\ 3}\hat{g}_2^{\ 1} - \hat{g}_1^{\ 1}\hat{g}_2^{\ 3}\right) \\ \bar{g}^3{}_3 = G^{1/2}\left(\hat{g}_1^{\ 1}\hat{g}_2^{\ 2} - \hat{g}_1^{\ 2}\hat{g}_2^{\ 1}\right) \end{gathered} \tag{A.2b}$$

$$\begin{aligned} g^{1/2} = {} & G^{1/2}\hat{g}_3^{\ 1}\left(\hat{g}_1^{\ 2}\hat{g}_2^{\ 3} - \hat{g}_1^{\ 3}\hat{g}_2^{\ 2}\right) + G^{1/2}\hat{g}_3^{\ 2}\left(\hat{g}_1^{\ 3}\hat{g}_2^{\ 1} - \hat{g}_1^{\ 1}\hat{g}_2^{\ 3}\right) + \\ & + G^{1/2}\hat{g}_3^{\ 3}\left(\hat{g}_1^{\ 1}\hat{g}_2^{\ 2} - \hat{g}_1^{\ 2}\hat{g}_2^{\ 1}\right) \end{aligned}$$

We recall standard definition of deformation gradient $\mathbf{F}$, displacement field $\mathbf{u}$, displacement gradient $\mathbf{H}$ and strain tensor $\mathbf{E}$ (see for example [2])

$$\mathbf{F} = \partial \mathbf{x} / \partial \mathbf{X} \ , \ \mathbf{u} = \mathbf{x} - \mathbf{X} \ , \ \mathbf{H} = \partial \mathbf{u} / \partial \mathbf{X}$$
$$\mathbf{E} = \frac{1}{2}(\mathbf{H} + \mathbf{H}^{\mathrm{T}}) \tag{A.3}$$

where $(^{\mathrm{T}})$ denotes right transpose. Using (A.1-A.3) we derive approximation for Caushy's strain tensors,

$$\mathbf{H}^{h} = \sum_{k=1}^{3}\sum_{i=1}^{3} \hat{H}^{k}{}_{i}(\mathbf{G}_{k} \otimes \mathbf{G}^{i}) \ , \ \hat{H}^{k}{}_{i} = \hat{g}_{i}{}^{k} - \delta^{k}{}_{i} \ , \ (i,k = 1,2,3)$$
$$\mathbf{E}^{h} = \sum_{m=1}^{3}\sum_{n=1}^{3} \hat{E}^{m}{}_{n}(\mathbf{G}_{m} \otimes \mathbf{G}^{n}) \tag{A.4}$$
$$\hat{E}^{m}{}_{n} = \frac{1}{2}\left( \hat{H}^{m}{}_{n} + \sum_{i=1}^{3}\sum_{k=1}^{3} \hat{H}^{k}{}_{i} G^{im} G_{kn} \right) , \ (m,n = 1,2,3)$$

where $\otimes$ denotes tensor product. Using constitutive law (7b), approximation for Coushy's stress tensor is derived

$$\boldsymbol{\sigma}^{h} = \sum_{m=1}^{3}\sum_{n=1}^{3} \hat{\sigma}^{m}{}_{n}(\mathbf{G}_{m} \otimes \mathbf{G}^{n})$$
$$\hat{\sigma}^{m}{}_{n} = \lambda(\sum_{i=1}^{3} \hat{E}^{i}{}_{i})\delta^{m}{}_{n} + 2\mu \hat{E}^{m}{}_{n} \ , \ (m,n = 1,2,3) \tag{A.5}$$

kinematic quantity $\mathbf{B}^{i}$ detailed by

$$\mathbf{B}^{i} = \sum_{k=1}^{3} \overline{B}^{i}{}_{k} \mathbf{G}^{k} \ , \ \overline{B}^{i}{}_{k} = g^{1/2} \sum_{m=1}^{3} N^{i}{}_{,m} \overline{g}^{m}{}_{k} \ , \ (i = 0,..,9) \tag{A.6}$$